\documentclass[12pt,showpacs]{revtex4}

\usepackage[english]{babel}
\usepackage{pst-node}
\usepackage[dvips]{graphicx}

\newcommand{\be}{\begin{eqnarray}}
\newcommand{\ee}{\end{eqnarray}}
\newcommand{\ba}{\begin{array}}
\newcommand{\ea}{\end{array}}
\newcommand{\pa}[1]{\left(#1\right)}
\newcommand{\paq}[1]{\left[#1\right]}

\newcommand{\ds}{\displaystyle}

\begin{document}

\title{Event trigger generator for resonant spherical detectors of 
gravitational waves}
\author{Stefano Foffa and Riccardo Sturani}
\affiliation{D\'epartment de Physique Th\'eorique, Universit\'e de Gen\`eve, 
Geneva, Switzerland\\ 
stefano.foffa, riccardo.sturani@.unige.ch}

\begin{abstract}
We have set up and tested a pipeline for processing the data from a 
spherical gravitational wave detector with six transducers. 
The algorithm exploits the multichannel capability of the system and
provides a list of candidate events with their arrival direction.
The analysis starts with the conversion of the six detector outputs into the 
scalar and the five quadrupolar modes of the sphere, which are proportional to 
the corresponding gravitational wave spherical components.
Event triggers are then generated by an adaptation of the
WaveBurst algorithm.
Event validation and direction reconstruction are made by cross-checking
two methods of different inspiration: geometrical (lowest eigenvalue) and
probabilistic (maximum likelihood). The combination of the two methods
is able to keep substantially unaltered the efficiency and can reduce 
drastically the detections of fake events (to less than ten per cent). We show 
a quantitative test of these ideas by simulating the operation of the resonant 
spherical detector miniGRAIL, whose planned sensitivity in its frequency band 
(few hundred Hertz's around $\rm{3 kHz}$) is comparable with the present LIGO
one.
\end{abstract}
\maketitle

\section{Introduction}
What makes a resonant spherical detector \cite{miniG,Aguiar:2006va} really 
different from bars \cite{bar_web} and interferometers \cite{intf_web} is the 
fact of being a multichannel detector.
As result, a sphere has almost isotropic sensitivity and enables
to reconstruct the gravitational wave (GW) direction. 
After the seminal work of Wagoner and Paik \cite{Wagoner}, where the basic 
features of the detector have been pointed out, more and more detailed sphere 
models and configurations have been studied by several authors 
\cite{Zhou:1995en,Coccia:1995yi,Lobo:1996ae,Stevenson:1996rw,Merkowitz:1997qc,Merkowitz:1997qs,Lobo:2000hy,Gasparini:2006vb,Gottardi:2006gn}.
In particular, various solutions have been proposed to the problems of the 
ideal transducers configuration \cite{Lobo:2000hy}, parameters reconstruction 
\cite{Merkowitz:1997qs,Zhou:1995en}, and multidimensional data analysis 
\cite{Stevenson:1996rw}.

The analysis method that we have set up and studied in the present work is the 
result of a synthesis and a refinement of some of these contributions, and is 
intended to represent the core of the pipeline that we are building for the 
miniGRAIL detector.
MiniGRAIL is one of the two small prototypes of resonant spherical
detectors which will be soon operating in the $3$kHz region,
with the possibility of detecting quasi-normal modes of strange core neutron
stars \cite{Benhar:2006ip} or burst from sub-solar mass black hole
coalescences \cite{Nakamura:1997sm}.

A greater variety of sources could be studied by building a larger sphere
operating in the sub-kHz region and our method, being completely general,
could be applied to such detectors as well.

Sec.~\ref{sec:model} of the paper is devoted to building an accurate detector 
model 
and to the generation of a simulated set of data. We concentrate on the case 
of six transducers placed in a specific configuration (the so-called TIGA
\cite{Merkowitz:1997qc}),
which best preserves the spherical symmetry of the
detector, and has been adopted by the miniGRAIL team.\\
For the same reason as above, the parameters of the numerical sphere
model (mass, radius, Poisson's ratio, modes and transducers frequencies, 
temperature, quality factors, readout electronic components) have been chosen 
in order to mimic the features of miniGRAIL, including asymmetries and
possible ``imperfections''.

We work out the spectrum of
each transducer's output and we generate simulated data starting from 
the elementary noises which are involved at different stages in the detector. 
Then, we dicuss the problem of extracting the GW parameters
out of the data.

In sec.~\ref{sec:trigger} we describe the method to generate trigger of
events.
 We used WaveBurst, the event trigger generator used by the Burst working 
group of the LIGO-VIRGO joint collaboration, suitably adapted for our 
multi-mode analysis.
Once an event trigger is obtained, the five quadrupolar modes provide a 
redundant description of a 
GW signal, which depends on four parameters (the amplitudes of the two
polarizations and the two angles identifying the source direction). We
exploit this redundancy to distinguish true GW signals from other non-Gaussian
noise (like glitches, for instance).
The multimode analysis is then able to compute all the relevant quantities 
of a GW and to reduce the false alarm rate.

 Sec.~\ref{sec:result} contains some results of the analysis, showing the 
level of signal-to-noise ratio required to obtain the desired accuracy in the 
direction reconstruction and efficency.

\section{Modeling the detector}
\label{sec:model}

\subsection{Equations for modes, transducers and readout currents}
\begin{figure}
\centering
\includegraphics[width=.8\linewidth]{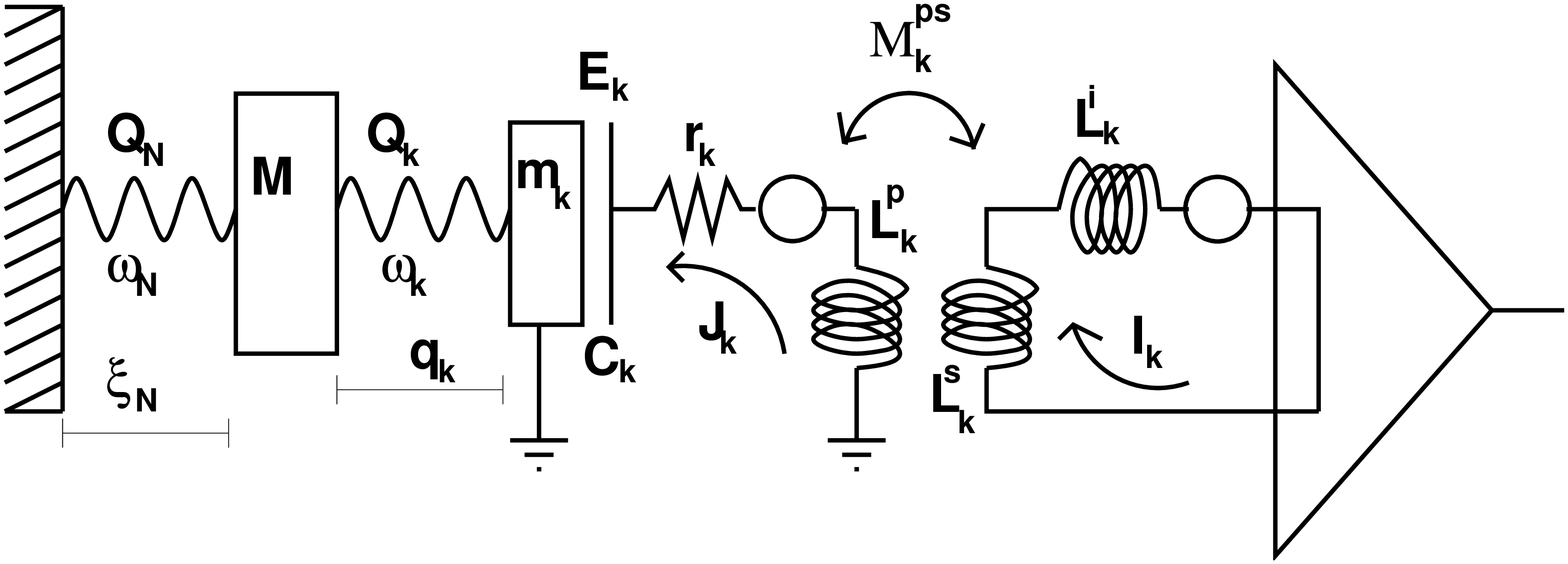}
\caption{The readout scheme of one of the tranducers. The surface of the
 sphere is modeled as an
  oscillating mass $M$, whose displacement is amplified by the six trasducers
of mass $m_k$, $k=1..6$, placed at different positions on the sphere. For each
transducer an electric circuit (primary) converts the mechanical displacement 
into a current $I_k$ which is then converted into a larger current $I_k$ in
  the secondary circuit. A SQUID finally amplifies the secondary current.}
\label{readout}
\end{figure}

The sphere is well modeled \cite{Merkowitz:1997qc,Gottardi:2006gn} as a set of
coupled mechanical and electric oscillators,
describing the dynamics of the relevant sphere vibrational modes,
of the transducers, and of the electrical circuit that are at the core of the 
readout devices, see fig.~\ref{readout}.\\
As the relevant equations have been already discussed in detail by several
authors  (see for instance the references above),
we jump to the mathematical core of the problem, skipping 
introductory material and definitions that can be found in the literature.

In Fourier space, the detector is described by a system
of linear and algebraic equations
\begin{eqnarray}
\label{findIk}    
\{\xi_N, q_k, J_k, I_k\}= {\cal Z}\left[f_N, f^t_k, f^p_k, f^s_k,f^I_k\right]\,,
\end{eqnarray}
where the $\xi_N$ are the amplitudes of the various radial modes of the sphere,
the $I_k,J_k$ are the currents flowing in the $k-$th transducer
electric circuit of the readout, the $q_k$ the positions of the 
relative mechanical oscillators, and the $f$'s the stochastic forces
related to the various dissipative components of the detector.
As to the operator ${\cal Z}$, its exact form can be deduced from the
equations contained in \cite{Merkowitz:1997qc,Gottardi:2006gn}:
for our purposes, what matters is that this operator can be computed
numerically as a function of the detector parameters.

\begin{figure}
\centering
\includegraphics[width=.7\linewidth]{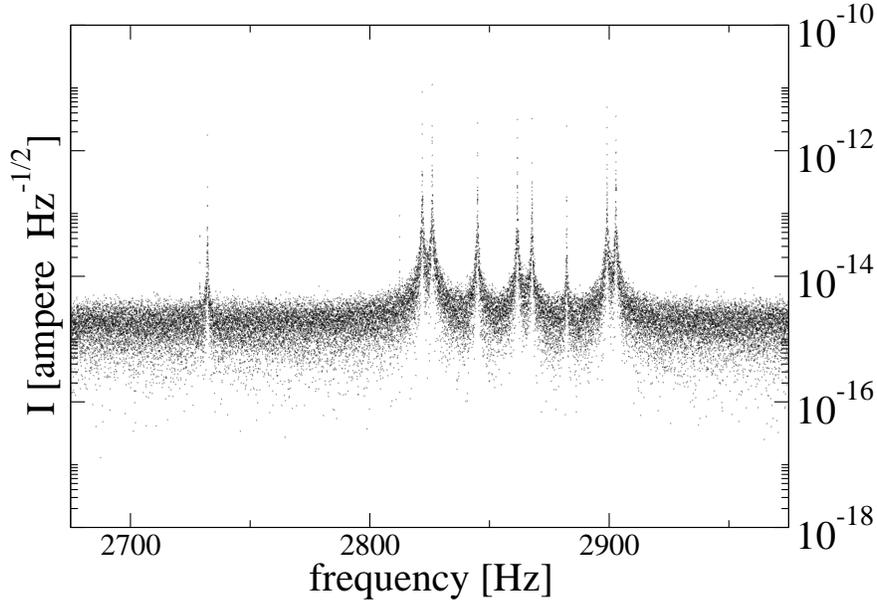}
\caption{The simulated output of transducer \#0.}\label{tout0_rec}
\end{figure}

Since the noise spectral densities of the stochastic forces $f$'s are known
(again, see \cite{Merkowitz:1997qc,Gottardi:2006gn}), a possible detector
configuration can be produced thanks to a random number generator; then
the corresponding output currents spectral
densities can be derived through eq.(\ref{findIk}),
as shown in figure \ref{tout0_rec}.

\subsection{From currents to the GW wave modes}
When a GW impinges the detector, a deterministic force is added 
to the quadrupolar components of $f_N$:
\begin{eqnarray}\label{newredfN}
f_N\rightarrow f_N+\frac{1}{2}R\chi_N\omega^2 h_N\quad{\rm for}\quad 
N=0,\dots,4\, ,
\end{eqnarray}
\begin{eqnarray}\label{newhN}
h_{ij}=\sum_{N=0,4}{\cal Y}^2_{m_N, ij} h_N\, ,
\end{eqnarray}
where $R$ is the sphere radius and
$Y^2_{m_N}=\sum_{i,j}{\cal Y}^2_{m_N, ij}n^i n^j$, being $n^i$ the versor
of the arrival direction of the wave and $m_N=0,1c,1s,2c,2s$, according to
the conventions used in \cite{Zhou:1995en}.

When six transducers are present, as in the TIGA configuration,
the system is in principle overconstrained as the $5$ 
quadrupolar $h_M$'s are to be determined by the $6$ outputs.

In the idealized case, the solution is provided by the {\it mode channels} ,
i.e. 5 linear combinations of the currents which are in one to one
correspondence with the sphere quadrupolar modes \cite{Merkowitz:1997qc},
while the sixth, remaining combination is insensitive to any such modes.

In the presence of noise and of asymmetries however the exact definition
of mode channels is not trivial and require the addition of a sixth
vibrational mode of the sphere in the problem, in order to make the system
invertible \cite{Foffa:2008pm}.

We chose then to include the scalar mode in the system,
as this is the mode that in the idealized case is probed by the sixth
linear combination of current outputs.
Moreover, the determination of the scalar mode allows, in principle,
to test alternative theories of gravity (where scalar GW's can exist),
or to build a veto, as events exciting such mode cannot be 
reduced to Einstein general relativity GW's.  

Given these premises, the problem can now be numerically solved
(for every value of $\omega$) to give 
\begin{eqnarray}\label{transf}
h_N(\omega)= T_{Nk}(\omega)\cdot I_k(\omega)\, ,
\end{eqnarray}
where $T_{Nk}(\omega)$ defines the  {\it transfer function} (which in this 
case is actually a transfer matrix) of the system.

\begin{figure}
\centering
\includegraphics[width=.7\linewidth]{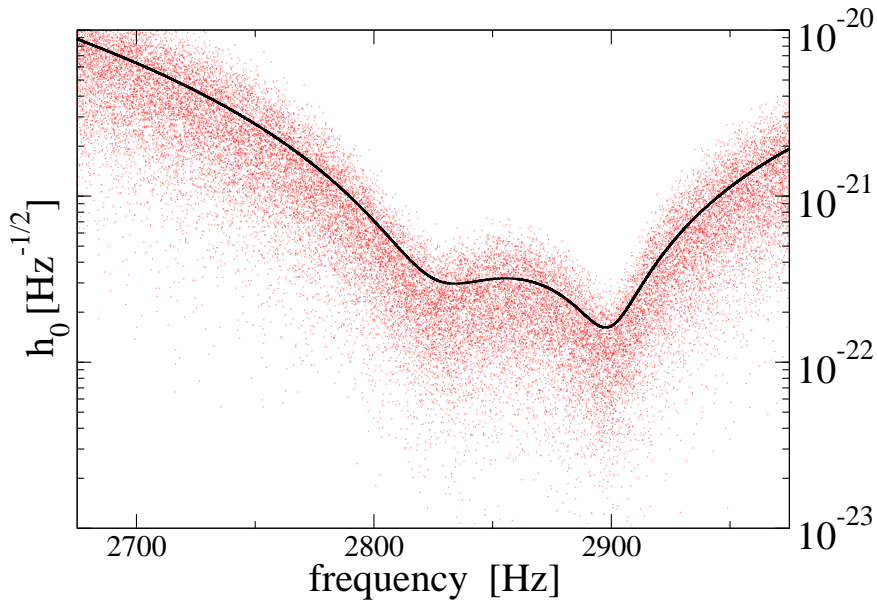}
\caption{The mode $h_0$ as it has been reconstructed starting from an actual
noise realization, superimposed with its expectation value.}\label{hout0_rec}
\end{figure}

We can now use eq.(\ref{transf}) to express any realization of the detector
output in terms of GW quadrupolar modes, as shown in figure \ref{hout0_rec}.
Moreover eq.(\ref{transf}) can be combined with the results of
the previous paragraph in order to give the $h_N$ spectral density matrix:
\begin{eqnarray}\label{noiseout}    
<I_k(\omega) I^*_{k'}(\omega')>\rightarrow T_{Nk}
\rightarrow<h_N(\omega) h^*_{N'}(\omega')>\equiv 
S^h_{NN'}(\omega)\delta(\omega-\omega')\, ,
\end{eqnarray}
which can be used to estimate the detector sensitivity.

\section{The analysis pipeline}
\label{sec:trigger}
We have just hinted how to generate six time series
corresponding to the quadrupolar plus scalar mode.
As at this stage we are not interested in the scalar GW
component, then we will restrict 
the analysis to the five quadrupolar modes, which will be enough and necessary 
to reconstruct the most general \emph{symmetric, traceless} $3\times3$ matrix.

Still this is a redundant description of a GW, and we will exploit this 
redundancy to discriminate between real GW signals and excitations of the
modes due to disturbances other than gravitational.

\subsection{The scalar trigger}
If one knew exactly the form of the expected signal, the optimal strategy would
be to perform a multidimensional matched filtering, but we would like to find 
triggers out of the stretch of data, without any need of a
detailed knowledge of the signal.
This can be done by building the following quantity:
\be \label{hmode}
H(\omega) = (h\cdot h)_S\, ,
\ee
which is a generalization of the one proposed in \cite{Gasparini:2006vb},
and takes account of the differences and of the correlations among
the various quadrupolar modes, as the r.h.s. is defined by
\be
(X\cdot Y)_S\equiv \sum_{N,N'=0\dots4}X^\dag_N(\omega)
(S^h_{NN'})^{-1}(\omega)Y_{N'}(\omega)\,.
\ee
The Fourier transform of this quantity is then fed to a suitably
adapted version of WaveBurst,
one of the burst event trigger generators used in the LIGO data 
analysis \cite{Klim,Klim2}.
The WaveBurst algorithm make use of the wavelet decomposition, and among 
the bank of wavelet packets we picked the Symlet base with filter length sixty 
\cite{Mal}.
Using an orthogonal wavelet transformation the time series are converted into 
wavelet series $W_{ij}$ where $i$ is the time index and $j$ the wavelet 
\emph{layer} index. Each wavelet layer can be associated with a certain 
frequency band of the initial time series.
The time frequency resolution varies according to the \emph{decomposition 
level}, if ${\cal N}$ is the number of data in the initial time series, at 
decomposition level $n$ the number of layers is $2^n$, each with 
${\cal N}\times 2^{-n}$ points (the procedure can go on as long as 
${\cal N}2^{-n}$ is an integer).

For each layer a fixed fraction of the coefficients $W_{ij}'s$ with the 
largest absolute magnitude is selected and WaveBurst keeps as triggers only 
those coefficients whose selection is robust under change of the decomposition 
level. Each trigger consists of highlighting a connected region in the 
time-frequency plane where the wavelet coefficients exceed an adaptive 
threshold (e.g. the 0.5\% higher coefficients).

An eventual GW signal impinging on the detector will have different strength
in different channels, depending on the arrival direction and polarization, 
whereas in $H$, see eq.(\ref{hmode}), it will imprint almost exactly the same 
signal, no matter its polarization nor arrival direction.

Once the trigger has been established the analysis is performed on the modes 
$h_N$'s, by collecting the values of the wavelet coefficients for each
channel and for each trigger. At this point we can try to reconstruct the 
arrival direction in (at least) two different ways.
 
As a first method one hand one can find the values of $\theta,\phi$ which 
maximize the \emph{a posteriori} probability of having our stretch of data 
given that a GW hit the detector see sec.\ref{likel}.
However this method will give a determination of the would be GW direction no 
matter if a real GW has excited the detector or a glitch, say, or
any other noise excitation not compatible with a transverse, traceless GW 
event has taken place.
To confirm this direction determination we combine it with a different, 
geometric method, see sec.~\ref{geom}. When the two methods do not determine 
the same directions, within some tolerance to be discussed quantitatively in 
sec.~\ref{sec:result}, the event can be discarded as spurious.

\subsection{Likelihood method for direction reconstruction}
\label{likel}
The posterior probability of having a given stretch of data $\{h_N\}$ 
assuming a GW with 
polarization strength $h_+$ and $h_\times$ from the direction identified by the
usual polar angles $\theta,\phi$ is denoted by 
$p(\{h_N\}|h_+,h_\times,\theta,\phi)$. 
The $N$ mode response to the gravity wave is
\be
\xi_N=F_N^{+}(\theta,\phi)h_{+}+ F_N^{\times}(\theta,\phi)h_{\times}\,,
\ee
where $F_N^{+,\times}$ is the pattern function of the $N$ mode for the 
$+,\times$ polarization.

Following the standard procedure \cite{Flanagan:1997kp,Anderson:2000yy}, 
the following likelihood ratio can be defined
\be
\Lambda=\frac{p(\{h_N\}|H_{\xi_N})}{p(\{h_N\}|H_0)}\,,
\ee
where the $H_{\xi_N}$ is the hypothesis that a GW characterized by $\xi_N$
is present in the data and $H_0$ is the hypothesis that no GW is in the data.

For stationary, Gaussian, white noise with zero mean, and by taking into
account that the noises in the different channels are correlated,
the above mentioned probability densities are 
\be
\ba{rcl}
\ds
p(\{h_N\}|H_{\xi_N})&\propto &\exp\paq{-((h-\xi)\cdot(h- \xi))_S/2}\,,\\
\ds
p(\{h_N\}|H_0)&\propto &\exp\paq{-(h\cdot h)_S/2}\,,\\
\ea
\ee
and the logarithm of the likelihood ratio can then be expressed as 
\be \label{loglike}
\mathcal L=\ln\pa{\Lambda}={\cal R}{\rm e} \paq{(\xi\cdot h)_S}
-\frac{1}{2}(\xi\cdot\xi)_S\,.
\ee
By maximizing eq.~(\ref{loglike}) with respect to $h_+,h_\times$
a function $L(\theta,\phi)$ of the angles alone is found
\be
\ba{rl}
\ds
L(\theta,\phi)=\frac{1}{2}\left[(F^+\cdot F^+)_S(F^{\times}\cdot F^{\times})_S
-(F^+\cdot F^{\times})^2_S\right]^{-1}\times&\\
\ds
\!\!\!\!\!\!\!\!\!\!\!\!\!\!\!\!\!\!\!\!\!
\left[(F^{\times}\cdot F^{\times})_S (h\cdot F^+)^2_S +
(F^+\cdot F ^+)_S(h\cdot F^{\times})^2_S-
2(h\cdot F^+)_S(h\cdot F^{\times})_S(F^+\cdot F^{\times})_S\right]&
\ea
\ee
The values of $\theta,\phi$ (which enter the expression for 
the $F_N$'s ) maximizing $\sum_{\{trigger\}} L(\theta,\phi)$,
that is the sum of the likelihood over every point of  time-frequency
plane exceeding the threshold, give the arrival direction of the candidate
event.

\subsection{Geometric method for direction reconstruction}
\label{geom}
Another method to reconstruct the direction of arrival of the GW by exploiting 
the (redundancy of the) five quadrupolar modes is based on linear algebra 
considerations \cite{Merkowitz:1997qc}.
A general metric perturbation $h_{ij}$ (the temporal components are suppressed 
here) representing a GW coming from the $z$ axis can be written as 
\be \label{hij_ideal}
h_{ij}=\left(\ba{cccc}
h_+ & h_\times & 0\\
h_\times & -h_+ & 0\\
0 & 0 & 0\\
\ea\right)\,.
\ee
For a different incoming direction the specific shape of the metric 
perturbation will be different, but the \emph{eigenvalues} of the matrices 
will be the same, thus implying that a generic GW is always represented by 
a \emph{traceless} matrix with a \emph{zero eigenvalue} 
(thus implying that the non zero eigenvalues are equal in magnitude and 
opposite in sign). In particular the direction of the zero eigenvalue is the 
propagation direction of the GW. 
Of course none of the eigenvalues is expected to be exactly zero at every 
instant within the duration of the trigger, so for each instant of time the 
direction of the smallest absolute magnitude eigenvalue is computed.
We then weight each direction with an empirical factor $1/r$, taking into 
account how the eigenvalues are close to the ideal, noiseless situation 
described by (\ref{hij_ideal}).

After ordering the three eigenvalues $\lambda_i$ so that 
$|\lambda_0|\leq |\lambda_1|\leq |\lambda_2|$ we define the following
quantities
\be
r\equiv\frac{\sqrt 2|\lambda_0|}{\sqrt{\lambda_1^2+\lambda_2^2}}\,.
\ee
For a perfect GW like (\ref{hij_ideal}) $r$ vanishes, thus the 
smallest it is, the less the noise is contaminating the GW signal.

\begin{figure}
\centering
\includegraphics[width=.7\linewidth]{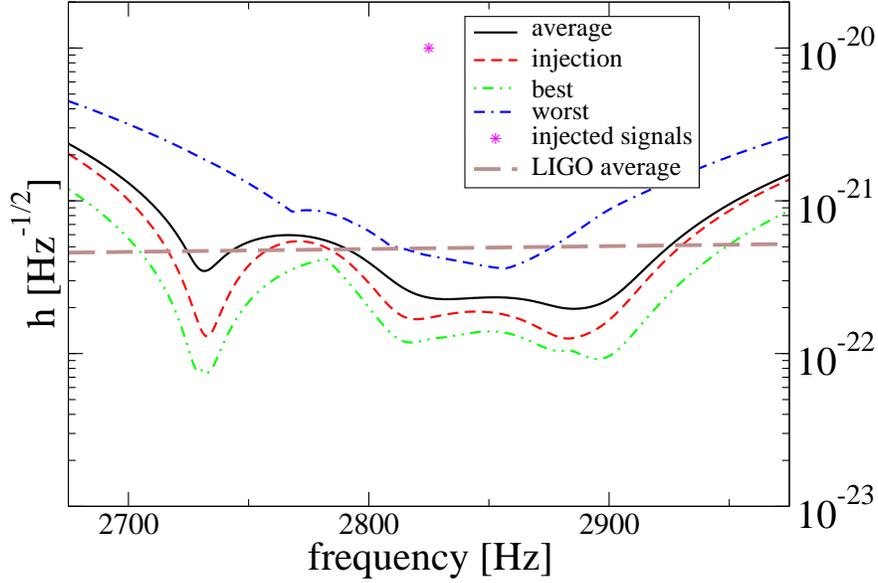}
\caption{The stars indicate the injections, while the dashed line is the strain
sensitivity for the arrival direction of the injections.
For comparison, we have displayed also also the averaged sensitivity
(continuous line), as well as the ones corresponding, at any given
value of the frequency, to the best and worst possible directions.
Finally, the thick dashed line is the LIGO current averaged on the sky.
%\footnote{Strictly speaking, the sky average of an
%interferometer strain sensitivity is infinite because of the presence of
%a blind direction. For the sake of a comparison, we have simply divided the
%sensitivity for optimal direction and polarization by the square root
%of the angular efficiency factor $F$, which for an interferometer is $F=2/5$.}
}
\label{hout_inj}
\end{figure}

\section{Results}
\label{sec:result}
To check our method we injected in software both GW signals and signals 
exciting of equal strength on the different channels, corresponding to a 
metric perturbation of the type
\be \label{hij_glitch}
h_{ij}=\left(\ba{cccc}
(\sqrt{3}-1)/\sqrt{3} & 1 & 1 \\
1 & -(\sqrt 3+1)/\sqrt 3 & 1\\
1  & 1 & 2/\sqrt 3\\
\ea\right)\,,
\ee
which is incompatible with a GW.
We investigated how well we could recover the injections of both real 
and fake signals in each of the channels with different waveforms.

\begin{figure}
\centering
\includegraphics[width=.55\linewidth]{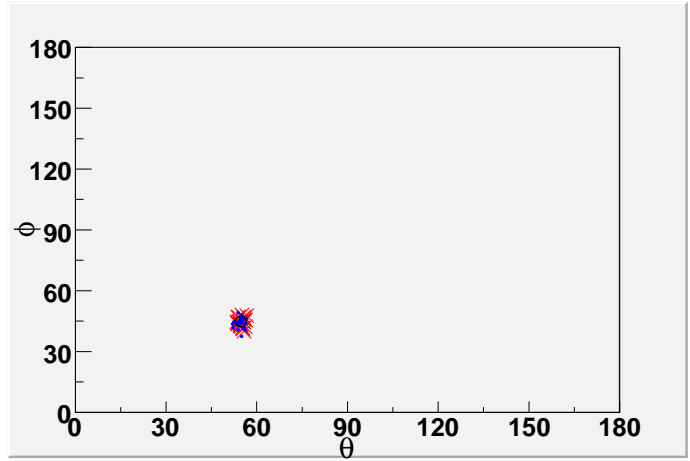}
\caption{The dots indicate direction reconstruction through the
  ``determinant'' method, the crosses through the likelihood one. The 
  injection direction is marked by a circle. Both dots and crosses fall very
  well on the top of the injection direction.}
\label{dir_inj_pl}
\end{figure}

\begin{figure}
\centering
\includegraphics[width=.55\linewidth]{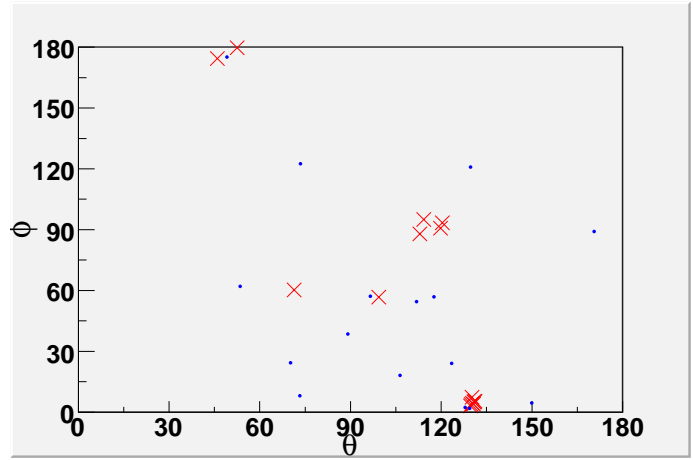}
\caption{The dots indicate direction reconstruction through the
  ``determinant'' method, the crosses through the likelihood one. The signal
  injected does not correspond to a GW.}
\label{dir_inj_fk}
\end{figure}

\begin{figure}
\centering
\includegraphics[width=.65\linewidth]{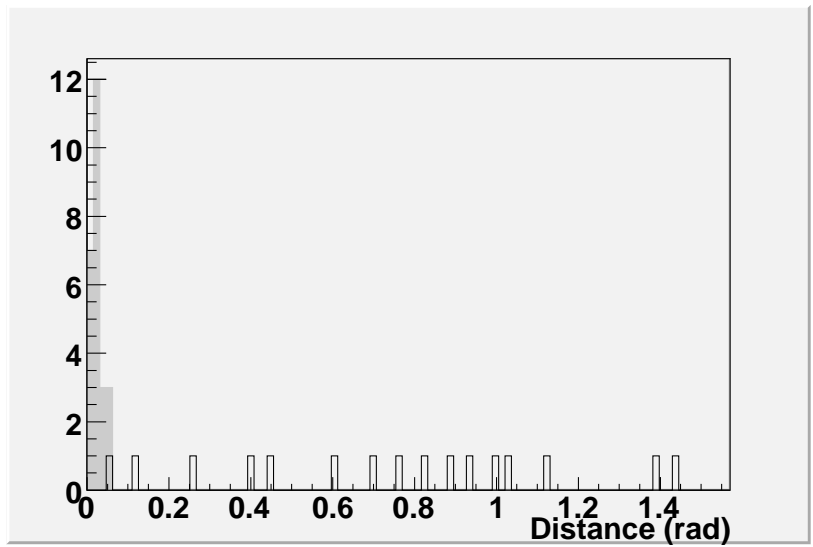}
\caption{Distribution of angular distances between the directions reconstructed
with the two different methods in the case of GW-injections (grey-filled) and 
fake-injections (black-transparent).}
\label{dist_plfc1}
\end{figure}

\begin{figure}
\centering
\includegraphics[width=.65\linewidth]{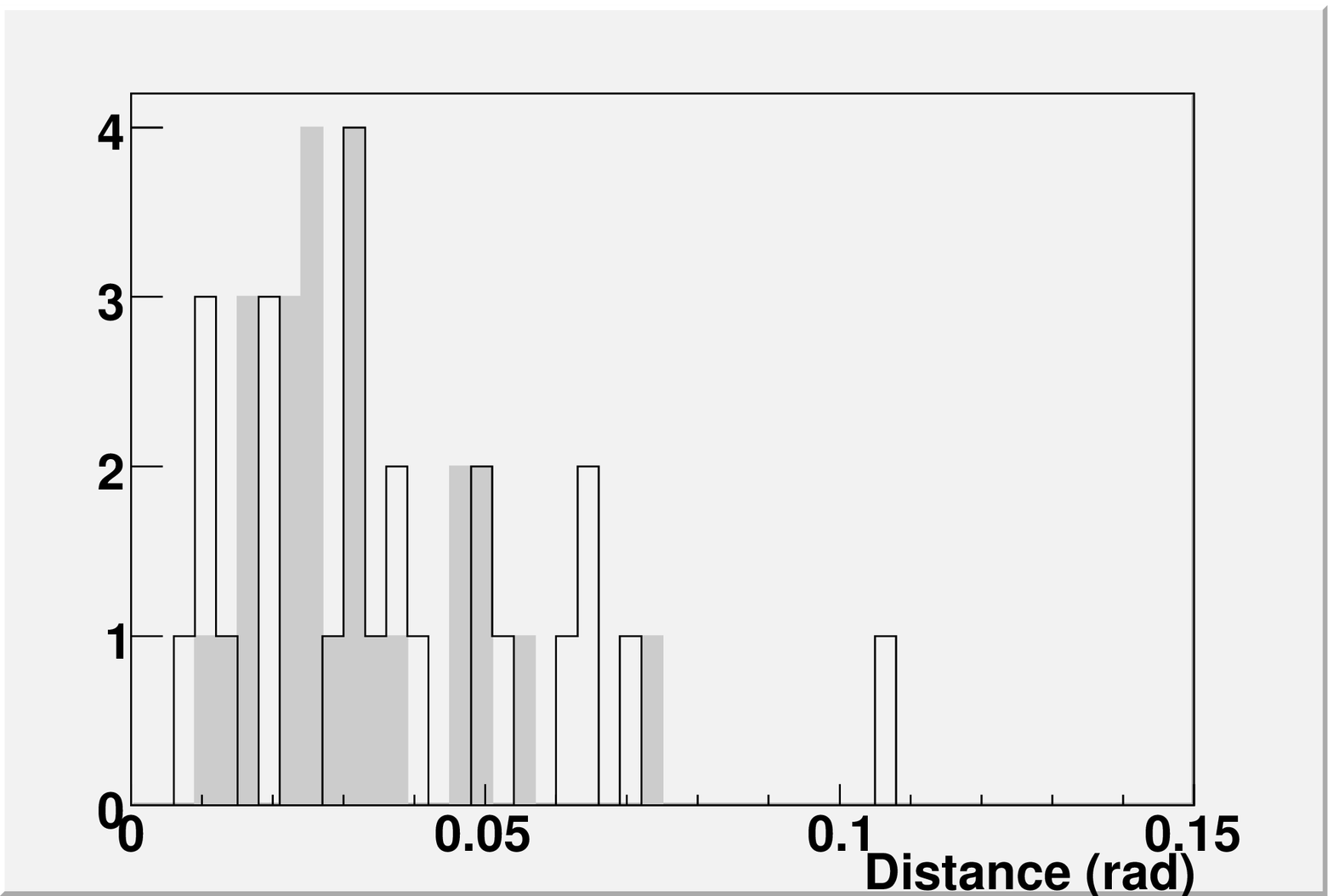}
\caption{Distribution of angular distances between 
the injection direction and the directions reconstructed with the likelihood 
(grey-filled) and with the determinant method (black-transparent) in the case 
of GW-injections. The average in the two cases is respectively 
$\mu_l=0.030$ ($\sigma=0.015$), $\mu_d=0.037$ ($\sigma=0.023$).}
\label{dist_plfc2}
\end{figure}

We injected sine-gaussian with a time-width of $50$msec, central 
frequency $2825Hz$  and with $h_{rss}=10^{-20}{\rm Hz}^{-1/2}$, 
see fig.~\ref{hout_inj}.
Fig.~\ref{dir_inj_pl} refers to an injection of 25 polarized GW-signals 
incoming from the direction specified by the angle $\theta=55^o,\phi=45^o$, 
whereas fig.~\ref{dir_inj_fk} refers to a signal of the type 
described by eq.(\ref{hij_glitch}), or fake-injection, with the same shape and 
strength.
The two direction identification methods agree much better in the case of GW
rather than in the fake-injection case see fig.~\ref{dist_plfc1}, where
an almost flat distribution of distances is obtained.
We then propose the combination of these methods can be used to reduce the
false alarm rate of a single detector.

Still, from the fig.\ref{dist_plfc1} it can be seen that by setting the
maximum acceptable distance to 0.1, say, one has perfect efficiency and 
still picks in some $8\%$ of the non GW-signals, thus suggesting that 
operating in coincidence with another detector is advisable.
Moreover fig.~\ref{dist_plfc2} hints that the likelihood method is more 
efficient than the determinant one in determining the direction.
\footnote{If instead the source direction is known, a match-filter can be 
applied.}

We thus conclude these display of preliminary results by claiming that the
information a spherical detector can provide can indeed determine the arrival
direction of a GW and can discriminate between real GW's and glitches, say,
even if more work is needed to make this result more quantitative and solid.

\section{Conclusion}
We have simulated a spherical detector and shown that a resonant sphere
detector is capable of detecting the direction of arrival of a gravitational 
wave. We have shown that at (amplitude) SNR$\simeq 53$ (corresponding to 
signals with $h_{rss}=1\times 10^{-20}{\rm Hz^{-1/2}}$) the likelihood method 
is well able
to determine the direction, but it is not able to discriminate between real GW
events and other kind of excitations, as for instance one with all modes
excited equally. By cross-checking this method with a different one like we did
it is possible to obtain an independent determination of the direction, which
leaves substantially unaltered the efficiency and lower the false alarm rate 
to 8\% of the rate of fake events.
More work has to be done to assess the method at different at SNR and with
higher statistics.

To further lower the false alarm rate it is advisable to work in coincidence, 
with another sphere, which could give an additional determination of the
arrival direction, or with interferometers, which are poor in direction
source determination, even if used as a network.

\section*{Acknowledgments}
The authors acknowledge support from the Boninchi foundation.
The authors wishes to thank Giorgio Frossati for his encouragement and support.
R.S. wishes to thank the organizers of the GWDAW-12 meeting.
During most of the time he has been working at this, R.S. has been supported by
an INFN post-doctoral grant. R.S. wishes to thank Gabriele Vedovato and Sergey 
Klimenko for their help with the WaveBurst algorithm.

\end{document}